\documentclass[preprint,superscriptaddress,showpacs,nofootinbib,aps]{revtex4}
 \usepackage{graphicx}
 \usepackage{bm}
 
\begin{document}
 \title{Study of $B_{c}$ ${\to}$ $J/{\psi}{\pi}$, ${\eta}_{c}{\pi}$
        decays with perturbative QCD approach}
 \author{Junfeng Sun}
 \affiliation{College of Physics and Information Engineering,
              Henan Normal University,
              Xinxiang 453007, China}
 \thanks{Mailing address}
 \affiliation{Theoretical Physics Center for Science Facilities,
              Chinese Academy of Sciences,
              P.O.Box 918(4),
              Beijing 100049, China}
 \author{Dongsheng Du}
 \affiliation{Institute of High Energy Physics,
              Chinese Academy of Sciences,
              P.O.Box 918(4),
              Beijing 100049, China}
 \author{Yueling Yang}
 \affiliation{College of Physics and Information Engineering,
              Henan Normal University,
              Xinxiang 453007, China}
 \begin{abstract}
 The $B_{c}$ ${\to}$ $J/{\psi}{\pi}$, ${\eta}_{c}{\pi}$ decays are
 studied with the perturbative QCD approach. It is found that
 the form factors $A_{0,1,2}^{B_{c}{\to}J/{\psi}}$ and
 $F_{0}^{B_{c}{\to}{\eta}_{c}}$ for the $B_{c}$ ${\to}$ $J/{\psi}$,
 ${\eta}_{c}$ transitions and the branching ratios are sensitive to the
 parameters ${\omega}$, $v$, $f_{J/{\psi}}$ and $f_{{\eta}_{c}}$,
 where ${\omega}$ and $v$ are the parameters of the charmonium wave
 functions for Coulomb potential and harmonic oscillator potential,
 respectively, $f_{J/{\psi}}$ and $f_{{\eta}_{c}}$ are the decay
 constants of the ${J/{\psi}}$ and ${{\eta}_{c}}$ mesons, respectively.
 The large branching ratios and the clear signals of the final states
 make the $B_{c}$ ${\to}$ $J/{\psi}{\pi}$, ${\eta}_{c}{\pi}$ decays
 to be the prospective channels for measurements at the hadron colliders.
 \end{abstract}
 \pacs{12.39.St  13.25.Hw}
 \maketitle

 \section{Introduction}
 \label{sec1}
 The Large Hadron Collider (LHC) is schedule to run in this year.
 At the era of the LHC, there is still a room for $B$ physics.
 The study of the decays of $B$ mesons is important and interesting
 for the determination of the flavor parameters of the
 Standard Model (SM), the exploration of $CP$ violation, the search
 of new physics beyond SM, and so on. The decays of $B_{u,d}$ mesons
 have been investigated widely by the detectors at the
 $e^{+}e^{-}$ colliders, such as the CLEO, Babar, Belle. The $B_{c}$
 meson could be produced abundantly and studied detailedly at the
 hadron colliders, such as the Tevatron and LHC. The study of the $B_{c}$
 mesons will highlight the advantages of $B$ physics.

 Compared with the $B_{u,d}$ mesons, the $B_{c}$ mesons have some
 special properties: (1) The $B_{c}$ mesons are the ``double
 heavy-flavored'' binding systems. We can study the two heavy flavors
 of both $b$ and $c$ quarks simultaneously with the $B_{c}$ mesons.
 (2) The $B_{c}$ mesons have much rich decay modes, because they have
 sufficiently large mass and that either $b$ or $c$ quarks can decay
 individually. The potential decays of the $B_{c}$ mesons permit us
 to over-constrain quantities determined by the $B_{u,d}$ meson decays.

 It is estimated that one could expect around $5$ ${\times}$ $10^{10}$
 $B_{c}$ events per year at LHC \cite{0412158}. The nonleptonic decays
 of the $B_{c}$ mesons have been studied in previous literature
 \cite{0412158,prd77p114004}.
 The theoretical status of the $B_{c}$ meson was reviewed in \cite{0412158}.
 In this paper, we will concentrate on the $B_{c}$ ${\to}$ $J/{\psi}{\pi}$,
 ${\eta}_{c}{\pi}$ decays using the perturbative QCD approach.
 There are several reasons :
 \begin{enumerate}
 \item[(i)] From the experimental point of view,
       the decay modes containing the signal of $J/{\psi}$ meson are among
       the most easily reconstructible $B_{c}$ decay modes, due to the
       narrow-peak of $J/{\psi}$ and the high purity $J/{\psi}$ ${\to}$
       ${\ell}^{+}{\ell}^{-}$. For example,
       the $B_{c}$ mesons are firstly discovered via $B_{c}$ ${\to}$
       $J/{\psi}{\ell}{\nu}$ by the CDF Collaboration in 1998 \cite{cdf98}.
       Recently the CDF and D0 Collaborations announced their accurate
       measurements on the $B_{c}$ mesons via $B_{c}$ ${\to}$ $J/{\psi}{\pi}$
       mode \cite{cdf07,d008}.
       Especially, compared with the semi-leptonic decays where the
       neutrino momentum is not detected directly, all final-state particles
       are detectable for the $B_{c}$ ${\to}$ $J/{\psi}{\pi}$,
       ${\eta}_{c}{\pi}$ decays.
       It is estimated that the ATLAS detector would be able to record about
       $5600$ events of $B_{c}$ ${\to}$ $J/{\psi}{\pi}$ per year \cite{0412158}.
       So $B_{c}$ ${\to}$ $J/{\psi}{\pi}$, ${\eta}_{c}{\pi}$
       decays may be two of the most prospective channels for measurements.
 \item[(ii)] From the phenomenological point of view:
       In recent years, several attractive methods have been proposed to
       study the nonleptonic $B$ decays,
       such as the QCD factorization \cite{9905312},
       perturbative QCD method (pQCD) \cite{9607214,9701233,0004004},
       soft and collinear effective theory \cite{prd63p114020,prd65p054022},
       and so on.
       The study of $B_{c}$ decays provides opportunities to test the $k_{T}$
       and collinear factorizations, to check the various treatments for the
       entanglement of different energy modes, to deepen our understanding on
       perturbative and nonperturbative contributions.
       These methods developed recently are widely applied to the nonleptonic
       two-body $B_{u,d,s}$ decays in literature, but with very few application
       of these methods on the $B_{c}$ meson decays.
       The appealing feature of the pQCD factorization \cite{9607214,9701233,0004004}
       is that form factors can be computed in terms of wave functions
       (nonperturbative contributions) and hard kernels (perturbative
       contributions arising from hard gluon exchange) assuming that
       additional soft contributions are suppressed by the Sudakov
       factor in the heavy quark limit.
       Although there is still some controversy about the pQCD method, for example,
       the problem of gauge invariant \cite{08070296}, the pQCD method has been
       extensively used in the past to study nonleptonic $B$ decays with fairly
       good phenomenological results \cite{pqcdlist}.
       In this paper, we will take the $B_{c}$ ${\to}$ $J/{\psi}{\pi}$,
       ${\eta}_{c}{\pi}$ decays as examples to discuss the $B_{c}$ decays in
       the perturbative QCD method.
 \item[(iii)] From the theoretical point of view:
       The $B_{c}$ ${\to}$ $J/{\psi}{\pi}$, ${\eta}_{c}{\pi}$ decays are similar
       to the $B_{q}$ ${\to}$ $D_{q}^{({\ast})}{\pi}$ (where $q$ $=$ $u$, $d$,
       $s$) decays with the ``spectator quark'' ansatz. The $B_{q}$ ${\to}$
       $D_{q}^{({\ast})}{\pi}$ decays have been studied with the pQCD method
       \cite{prd78p014018}. Compared with the $B_{q}$ ${\to}$ $D_{q}^{({\ast})}{\pi}$
       decays, the $B_{c}$ ${\to}$ $J/{\psi}{\pi}$, ${\eta}_{c}{\pi}$ decays
       are easy to deal with because that the $B_{c}$ meson and the $J/{\psi}$
       (or ${\eta}_{c}$) meson are heavy quarkonia and could be described
       approximatively by nonrelativistic dynamics.
       Given $m_{B_{c}}$ ${\simeq}$ $m_{b}$ $+$ $m_{c}$, the wave function of
       the $B_{c}$ mesons would be close to ${\delta}(x-m_{c}/{m_{B_{c}}})$ in
       the nonrelativistic limit (where $m_{B_{c}}$, $m_{b}$, and $m_{c}$ are
       the masses of the $B_{c}$ mesons, $b$ quark, and $c$ quark, respectively;
       $x$ denotes the momentum fraction of the $c$ quark in the $B_{c}$ meson).
       The wave functions for pion are well-defined.
       The only parameter is the wave function of the $J/{\psi}$ (or ${\eta}_{c}$)
       meson. So the $B_{c}$ ${\to}$ $J/{\psi}{\pi}$, ${\eta}_{c}{\pi}$ decays
       provide good platform to test quark potential models derived from QCD.
 \end{enumerate}

 This paper is organized as follows:
 In Section \ref{sec2}, we discuss the theoretical framework and compute the decay
 amplitudes for $B_{c}$ ${\to}$ $J/{\psi}{\pi}$, ${\eta}_{c}{\pi}$ with the
 perturbative QCD approach.
 The section \ref{sec3} is devoted to the numerical results.
 Finally, we summarize in Section \ref{sec4}.

 \section{Theoretical framework and the decay amplitudes}
 \label{sec2}

 \subsection{The effective Hamiltonian}
 \label{sec21}
 Using the operator product expansion and renormalization group (RG) equation,
 the low energy effective Hamiltonian for $B_{c}$ ${\to}$ $X_{c\bar{c}}{\pi}$
 decay can be written as
 (where $X_{c\bar{c}}$ $=$ $J/{\psi}$, ${\eta}_{c}$):
 \begin{equation}
 {\cal H}_{eff}\,=\,\frac{G_{F}}{\sqrt{2}} V_{cb}V_{ud}^{\ast} \Big\{
    C_{1}({\mu})Q_{1}+ C_{2}({\mu})Q_{2} \Big\} + \hbox{H.c.},
 \label{eq:hamiltonian}
 \end{equation}
 where $V_{cb}V_{ud}^{\ast}$ is the CKM factor accounting for the strengths
 of the concerned nonleptonic decay processes.
 The parameters $C_{i}({\mu})$ are Wilson coefficients which have
 been evaluated to the next-to-leading order with the perturbation theory.
 The expressions of the local operators are
 \begin{equation}
 Q_{1}=\big[\bar{c}_{\alpha}{\gamma}_{\mu}(1-{\gamma}_{5})b_{\alpha}\big]
       \big[\bar{d}_{\beta} {\gamma}^{\mu}(1-{\gamma}_{5})u_{\beta} \big],~~~~~
 Q_{2}=\big[\bar{c}_{\alpha}{\gamma}_{\mu}(1-{\gamma}_{5})b_{\beta} \big]
       \big[\bar{d}_{\beta} {\gamma}^{\mu}(1-{\gamma}_{5})u_{\alpha}\big],
 \label{eq:operator}
 \end{equation}
 where ${\alpha}$, ${\beta}$ are color indices.
 The essential problem obstructing the calculation of decay amplitude is how
 to evaluate the hadronic matrix elements of the local operators.

 \subsection{Hadronic matrix elements}
 \label{sec22}
 The calculation of the hadronic matrix elements is difficult due to the
 nonperturbative effects arising from the strong interactions.
 Phenomenologically, the simplest approach to hadronic matrix elements is the
 Bauer-Stech-Wirbel (BSW) model \cite{bsw} based on color transparency and
 naive factorization hypothesis, where the hadronic matrix elements are
 parameterized into the product of the decay constants and the transition
 form factors. One defect of the rough BSW method is that the hadronic matrix
 elements cannot cancel the renormalization scheme- and scale- dependence
 of the Wilson coefficients. To remedy this problem, the ``nonfactorizable''
 contributions must be taken into account.
 Using the Brodsky-Lepage approach \cite{prd22p2157}, the hadronic matrix
 elements can be written as the convolution of a hard-scattering amplitude,
 including some perturbative QCD contributions, and meson wave functions.

 Recently, a modified perturbative QCD formalism has been proposed under the
 $k_{T}$ factorization framework \cite{9607214,9701233,0004004}.
 The Sudakov effects are introduced to modify the endpoint behavior.
 The decay amplitudes are factorized into three convolution factors:
 the ``harder'' functions, the heavy quark decay subamplitudes, and
 the nonperturbative meson wave functions, which are characterized
 by the $W^{\pm}$ boson mass $m_{W}$, the typical scale $t$ of the
 decay processes, and the hadronic scale ${\Lambda}_{QCD}$, respectively.
 Using the resummation technique and the RG treatment,
 the final decay amplitudes can be expressed as
 \begin{equation}
 {\cal A}(B_{c}{\to}X_{c\bar{c}}{\pi})\, {\propto}\,
 C(t){\otimes}H(t){\otimes}{\Phi}_{B_{c}}(x_{1},b_{1}) {\otimes}
 {\Phi}_{X_{c\bar{c}}}(x_{2},b_{2}){\otimes}{\Phi}_{\pi}(x_{3},b_{3}),
 \label{eq:am01}
 \end{equation}
 where the Wilson coefficient $C(t)$ is calculated in perturbative theory
 at scale of $m_{W}$ and evolved down to the typical scale $t$ using the
 RG equations, ${\otimes}$ denotes the convolution over parton kinematic
 variables, $H(t)$ is the hard-scattering subamplitude, the wave functions
 ${\Phi}(x,b)$ absorb nonperturbative long-distance dynamics,
 $x$ is the longitudinal momentum fraction of the valence quark of the meson,
 $b$ is the conjugate variable of the transverse momentum of the valence
 quark of the meson. According the arguments in \cite{9607214,9701233,0004004},
 the amplitude of Eq.(\ref{eq:am01}) is free from the renormalization scale
 dependence.

 \subsection{Kinematic variables}
 \label{sec23}
\begin{figure}[ht]
 \includegraphics[250,570][400,770]{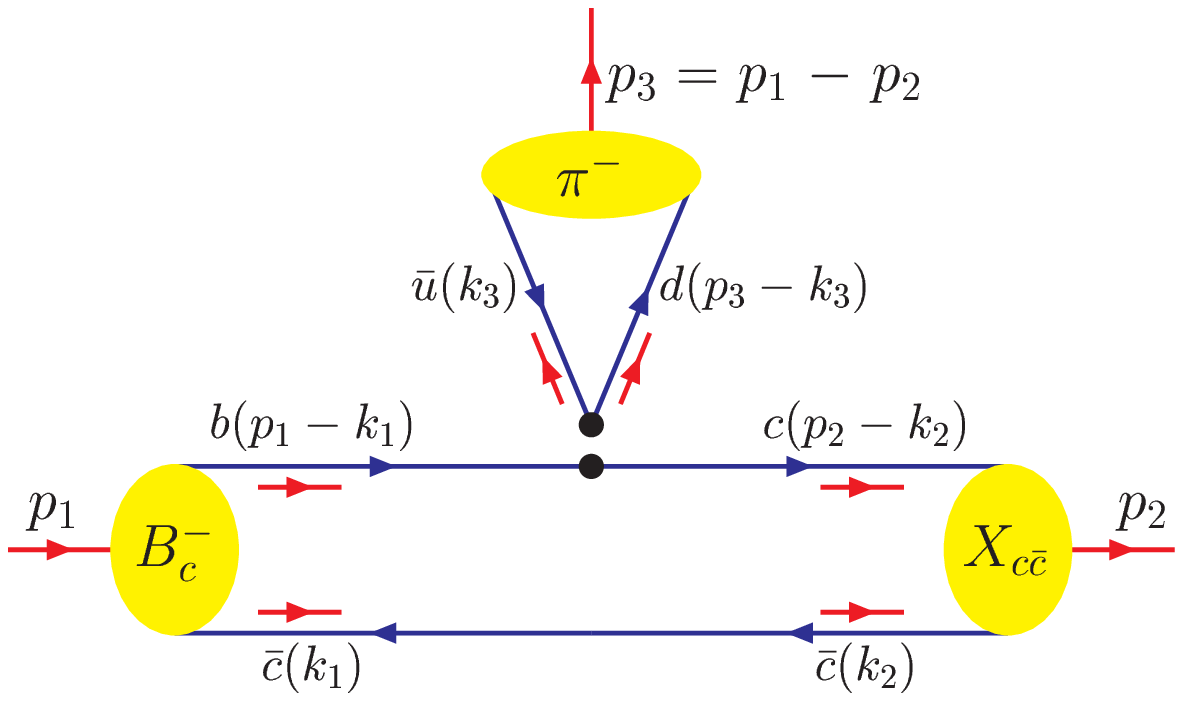}
 \caption{Kinematic variables for $B_{c}^{-}$ ${\to}$ $X_{c\bar{c}}{\pi}^{-}$ decays.}
 \label{fig1}
 \end{figure}

 For convenience, the kinematics variables are described in the terms of
 the light cone coordinate. The momenta of the valence quarks and hadrons
 in the rest frame of the $B_{c}$ meson are defined by
 \[ \begin{array}{lclcl}
    p_{1}=\frac{m_{B_{c}}}{\sqrt{2}}(1,1,\vec{0}_{\perp}), &~~&
    k_{1}=x_{1}p_{1}, &~~& n_{2}=(1,0,0), \\
    p_{2}=\frac{m_{B_{c}}}{\sqrt{2}}(1,r_{X_{c\bar{c}}}^{2},\vec{0}_{\perp}), &&
    k_{2}=x_{2}p_{2}+(0,0,\vec{k}_{2{\perp}}), & &
    {\epsilon}_{\parallel}=\frac{1}{\sqrt{2}r_{\psi}}(1,-r^{2}_{\psi},\vec{0}), \\
    p_{3}=\frac{m_{B_{c}}}{\sqrt{2}}(0,1-r_{X_{c\bar{c}}}^{2},\vec{0}_{\perp}), &&
    k_{3}=x_{3}p_{3}+(0,0,\vec{k}_{3{\perp}}), & & n_{3}=(0,1,0),
    \end{array} \]
 where the notation of momenta of $p_{i}$ and $k_{i}$ are displayed in FIG.\ref{fig1}.
 The null vectors $n_{2}$ and $n_{3}$ are the plus and minus directions,
 respectively. The mass of the ${\pi}$ meson is neglected. The momentum
 of the ${\pi}$ meson is chosen to be parallel to the null vector $n_{3}$.
 The mass ratios are $r_{X_{c\bar{c}}}$ $=$ $m_{X_{c\bar{c}}}/m_{B_{c}}$,
 $r_{b}$ $=$ $m_{b}/m_{B_{c}}$, $r_{c}$ $=$ $m_{c}/m_{B_{c}}$.

 \subsection{Bilinear operator matrix elements for mesons}
 \label{sec24}
 In terms of the notation in \cite{prd65p014007}, the nonlocal bilinear-quark
 operator matrix elements associated with the $B_{c}$ meson, ${\pi}$ meson,
 the longitudinally polarized $J/{\psi}$ meson, ${\eta}_{c}$ meson are
 decomposed into \cite{prd65p014007,prd71p114008}
 \begin{equation}
 {\langle}0{\vert}\bar{c}_{\alpha}(z)b_{\beta}(0){\vert}B_{c}^{-}(p_{1}){\rangle}
 =\frac{+i}{\sqrt{2N_{c}}}{\int}{\bf d}^{4}k_{1}\,{\rm e}^{-ik_{1}{\cdot}z}
  \Big[ \Big(\!\!\not{p}_{1}\!+\!m_{B_{c}} \Big) {\gamma}_{5} {\phi}_{B_{c}}(k_{1})
  \Big]_{{\beta}{\alpha}},
 \label{eq:wf-bc-01}
 \end{equation}
 \begin{eqnarray}
 & & {\langle}{\pi}^{-}(p_{3}){\vert}\bar{d}_{\alpha}(0)u_{\beta}(z){\vert}0{\rangle}
  \nonumber \\ &=& \frac{-i}{\sqrt{2N_{c}}}{\int}_{0}^{1}\!{\bf d}x_{3}\,
 {\rm e}^{+ix_{_{3}}p_{_{3}}z}\Big\{\!{\gamma}_{5}\Big[\!\not{p}_{3}{\phi}_{\pi}^{a}(x_{3})
 \!+\!{\mu}_{\pi}{\phi}_{\pi}^{p}(x_{3})\!-\!{\mu}_{\pi}(\!\not{n}_{3}\!\!\not{n}_{2}
 \!-\!n_{3}{\cdot}n_{2}){\phi}_{\pi}^{t}(x_{3})\Big]\Big\}_{{\beta}{\alpha}},
 \label{eq:wf-pion-01}
 \end{eqnarray}
 \begin{equation}
 {\langle}J/{\psi}(p_{2},{\epsilon}_{\parallel}){\vert}\bar{c}_{\alpha}(0)c_{\beta}(z)
 {\vert}0{\rangle} =\frac{1}{\sqrt{2N_{c}}}{\int}{\bf d}^{4}k_{2}\,
 {\rm e}^{+ik_{2}{\cdot}z}\!\!\not{\!\epsilon}_{\parallel}
  \Big[m_{J/{\psi}}{\phi}_{\psi}^{L}(k_{2})
 +\!\!\not{p}_{2}{\phi}_{\psi}^{t}(k_{2})\Big]_{{\beta}{\alpha}},
 \label{eq:wf-jpsi-01}
 \end{equation}
 \begin{equation}
 {\langle}{\eta}_{c}(p_{2}){\vert}\bar{c}_{\alpha}(0)c_{\beta}(z){\vert}0{\rangle}
 =\frac{-i}{\sqrt{2N_{c}}}{\int}{\bf d}^{4}k_{2}\,{\rm e}^{+ik_{2}{\cdot}z}
  \Big\{\!{\gamma}_{5}\Big[\!\not{p}_{2}{\phi}_{{\eta}_{c}}^{v}(k_{2})
  \!+\!m_{{\eta}_{c}}{\phi}_{{\eta}_{c}}^{s}(k_{2})\Big]\Big\}_{{\beta}{\alpha}},
 \label{eq:wf-etac-01}
 \end{equation}
 where the wave functions ${\phi}_{\pi}^{a}$, ${\phi}_{\psi}^{L}$,
 ${\phi}_{{\eta}_{c}}^{v}$ are twist-2,
 ${\phi}_{\pi}^{p}$, ${\phi}_{\pi}^{t}$, ${\phi}_{\psi}^{t}$,
 ${\phi}_{{\eta}_{c}}^{s}$ are twist-3,
 ${\mu}_{\pi}$ $=$ $m_{\pi}^{2}/(m_{u}+m_{d})$.
 Their expressions are collected in APPENDIX \ref{app01} and \ref{app02}.

 For the wave function ${\phi}_{B_{c}}$, we will take the nonrelativistic
 approximation as stated in the introduction, i.e.
 \begin{equation}
 {\phi}_{B_{c}}=\frac{f_{B_{c}}}{2\sqrt{2N_{c}}}{\delta}(x-r_{c}),
 \label{eq:wf-bc-02}
 \end{equation}
 where $N_{c}$ is the color number, $f_{B_{c}}$ is the decay
 constant of the $B_{c}$ meson.

 \subsection{$B_{c}$ ${\to}$ $X_{c\bar{c}}$ form factors}
 \label{sec25}
 The $B_{c}$ ${\to}$ $X_{c\bar{c}}$ form factors are defined as \cite{bsw,epjc28p515}:
 \begin{eqnarray}
  & &{\langle}{\eta}_{c}(p_{2}){\vert}\bar{c} {\gamma}^{\mu}b
     {\vert}B_{c}(p_{1}){\rangle} \nonumber \\ &=&
      \frac{m_{B_{c}}^{2}\!\!-\!m_{{\eta}_{c}}^{2}}{q^{2}}q^{\mu}
  F_{0}^{B_{c}{\to}{\eta}_{c}}(q^{2}) +
  \Big[\big(p_{1}+p_{2}\big)^{\mu}-\frac{m_{B_{c}}^{2}\!\!-\!m_{{\eta}_{c}}^{2}}{q^{2}}
  q^{\mu}\Big]F_{1}^{B_{c}{\to}{\eta}_{c}}(q^{2}),
 \label{eq:formfactor-01} \\
 & &{\langle}J/{\psi}(p_{2},{\epsilon}){\vert}\bar{c}{\gamma}^{\mu}{\gamma}_{5}b
    {\vert}B_{c}(p_{1}){\rangle} \nonumber \\ &=&
 +i\frac{({\epsilon}^{\ast}{\cdot}q)}{q^{2}}
  2m_{J/{\psi}}q^{\mu}A_{0}^{B_{c}{\to}J/{\psi}}(q^{2})
 +i{\epsilon}^{{\ast}{\mu}}(m_{B_{c}}\!\!+\!m_{J/{\psi}})A_{1}^{B_{c}{\to}J/{\psi}}(q^{2})
  \nonumber \\ & &
 -i\frac{({\epsilon}^{\ast}{\cdot}q)}{m_{B_{c}}\!\!+\!m_{J/{\psi}}}
  \big(p_{1}+p_{2}\big)^{\mu}A_{2}^{B_{c}{\to}J/{\psi}}(q^{2})
 -i\frac{({\epsilon}^{\ast}{\cdot}q)}{q^{2}}
  2m_{J/{\psi}}q^{\mu}A_{3}^{B_{c}{\to}J/{\psi}}(q^{2}),
 \label{eq:formfactor-02}
 \end{eqnarray}
 where $q$ $=$ $p_{1}$ $-$ $p_{2}$, ${\epsilon}^{\ast}$ denotes the polarization
 vector of the $J/{\psi}$ meson. $F_{0,1}^{B_{c}{\to}{\eta}_{c}}$ and
 $A_{0,1,2,3}^{B_{c}{\to}J/{\psi}}$ are the transition form factors.
 In addition, at large recoil limit, $q^{2}$ $=$ $0$, we have
 \begin{equation}
 F_{0}^{B_{c}{\to}{\eta}_{c}}(0)=F_{1}^{B_{c}{\to}{\eta}_{c}}(0),~~~~~~~~~~
 A_{0}^{B_{c}{\to}J/{\psi}}(0)=A_{3}^{B_{c}{\to}J/{\psi}}(0),
 \label{eq:formfactor-03}
 \end{equation}
 \begin{equation}
  A_{3}^{B_{c}{\to}J/{\psi}}(q^{2})
 =\frac{m_{B_{c}}\!\!+\!m_{J/{\psi}}}{2m_{J/{\psi}}}
  A_{1}^{B_{c}{\to}J/{\psi}}(q^{2})
 -\frac{m_{B_{c}}\!\!-\!m_{J/{\psi}}}{2m_{J/{\psi}}}
  A_{2}^{B_{c}{\to}J/{\psi}}(q^{2}).
 \label{eq:formfactor-04}
 \end{equation}

\begin{figure}[ht]
 \includegraphics[250,685][350,720]{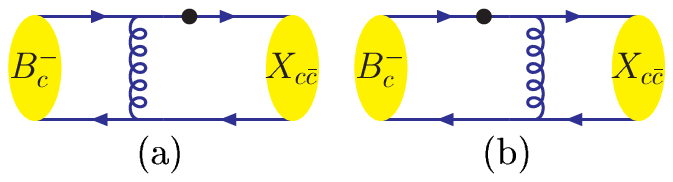}
 \caption{Feynman diagrams contributing to the $B_{c}$ ${\to}$ $X_{c\bar{c}}$ form factors.}
 \label{fig2}
 \end{figure}

 In the perturbative QCD approach, these form factors can be generally
 written as
 \begin{equation}
 A_{i}^{B_{c}{\to}J/{\psi}}~({\rm or}~F_{i}^{B_{c}{\to}{\eta}_{c}})~
 \, {\propto}~\, {\Phi}_{B_{c}}(x_{1},b_{1}) {\otimes}H(t){\otimes}
 {\Phi}_{X_{c\bar{c}}}(x_{2},b_{2}).
 \label{eq:formfactor-05}
 \end{equation}
 At large recoil region, the $B_{c}$ ${\to}$ $X_{c\bar{c}}$ transition
 is dominated by the single gluon exchange as depicted FIG.\ref{fig2}.
 The expressions for $F_{0}^{B_{c}{\to}{\eta}_{c}}$ and
 $A_{1,2}^{B_{c}{\to}J/{\psi}}$ are listed in APPENDIX \ref{app03}.

 \subsection{The decay amplitudes}
 \label{sec26}
 The $B_{c}$ ${\to}$ $J/{\psi}{\pi}$, ${\eta}_{c}{\pi}$ decays are tree
 dominated within the framework of Operator Product Expansion, and
 without pollution from penguins and annihilation diagrams.
 In the perturbative QCD approach, the Feynman diagrams are shown
 in FIG.\ref{fig3}, where (a) and (b) are factorizable topology,
 (c) and (d) are nonfactorizable topology.
 After a straightforward calculation using the modified perturbative
 QCD formalism Eq.(\ref{eq:am01}), we obtain the decay amplitudes
 \begin{equation}
 {\cal A}(B_{c}^{-}{\to}X_{c\bar{c}}{\pi}^{-})=
  \frac{G_{F}}{\sqrt{2}}V_{cb}V_{ud}^{\ast}\sum\limits_{i=a,b,c,d}
  {\cal A}_{\rm FIG.\ref{fig3}(i)},
 \end{equation}
 where the CKM matrix elements $V_{cb}V_{ud}^{\ast}$ $=$
 $A{\lambda}^{2}(1-{\lambda}^{2}/2-{\lambda}^{4}/8)$
 $+$ ${\cal O}({\lambda}^{8})$ with the Wolfenstein parameterization.
 The detailed expressions of ${\cal A}_{\rm FIG.\ref{fig3}(i)}$
 are shown in APPENDIX \ref{app04}. From the expressions, we can
 clearly see that only the twist-2 distribution amplitude of the
 ${\pi}$ meson contribute to the decay amplitudes.

 \section{Numerical results and discussions}
 \label{sec3}
 The branching ratios in the $B_{c}$ meson rest frame can be written as:
 \begin{equation}
  {\cal BR}(B_{c}{\to}X_{c\bar{c}}{\pi})= \frac{{\tau}_{B_{c}}}{8{\pi}}
   \frac{{\vert}p{\vert}}{m_{B_{c}}^{2}}
  {\vert}{\cal A}(B_{c}{\to}X_{c\bar{c}}{\pi}){\vert}^{2}
   \label{eq:br-01},
 \end{equation}
 where the common momentum ${\vert}p{\vert}$ $=$
 $(m_{B_{c}}^{2}-m_{X_{c\bar{c}}}^{2})/2m_{B_{c}}$,
 the lifetime and mass of the $B_{c}$ meson are
 $m_{B_{c}}$ $=$ $6.276$ ${\pm}$ $0.004$ GeV and
 ${\tau}_{B_{c}}$ $=$ $0.46{\pm}0.07$ ps \cite{pdg2008},
 respectively.
 Other input parameters are
 \[ \begin{array}{lll}
   m_{c}=1.5~\hbox{\rm GeV}, &
   m_{J/{\psi}}=3096.916{\pm}0.011~\hbox{\rm MeV~\cite{pdg2008}}, &
   f_{J/{\psi}}=405{\pm}14~\hbox{\rm MeV~\cite{prd71p114008}}, \\
   m_{b}=4.20^{+0.17}_{-0.07}~\hbox{\rm GeV~\cite{pdg2008}}, &
   m_{{\eta}_{c}}=2980.3{\pm}1.2~\hbox{\rm MeV~\cite{pdg2008}}, &
   f_{{\eta}_{c}}=420{\pm}50~\hbox{\rm MeV~\cite{prd71p114008}}, \\
   A=0.814^{+0.021}_{-0.022}~\hbox{\cite{pdg2008}}, &
   {\lambda}=0.2257^{+0.0009}_{-0.0010}~\hbox{\cite{pdg2008}}, &
   f_{B_{c}}=489{\pm}4~\hbox{\rm MeV~\cite{pos180}}.
   \end{array} \]
 If not specified explicitly, we shall take their central values as
 the default input.

 Our numerical results
 \footnote{Here, we think that theoretical prediction on input parameters,
 such as ${\omega}$ and $v$, relies on our educated guesswork. All values
 within allowed ranges should be treated on an equal footing, irrespective
 of how close they are from the edges of the allowed range. For example,
 we cannot say that the probability of ${\omega}$ $=$ $0.5$ GeV is less than
 that of ${\omega}$ $=$ $0.6$ GeV, while the error means the usual one
 standard deviation in the form of $A^{+{\delta}A}_{-{\delta}A}$ (such as
 the expression of $m_{B_{c}}$ $=$ $6.276$ ${\pm}$ $0.004$ GeV). So our
 numerical results had better to be given by a range to show the theoretical
 uncertainties, rather than the form of $A^{+{\delta}A}_{-{\delta}A}$.}
 on the form factors $F_{0}^{B_{c}{\to}{\eta}_{c}}$
 and $A_{0,1,2}^{B_{c}{\to}J/{\psi}}$ are listed in TABLE.\ref{tab1},
 where ${\omega}$ and $v$ are the parameters in the wave functions
 of Eqs.(\ref{eq:wf-jpsi-02})-(\ref{eq:wf-etac-03}) and
 Eqs.(\ref{eq:wf-jpsi-05})-(\ref{eq:wf-etac-06}), respectively.
 From the numbers in TABLE.\ref{tab1}, we can see that
 \begin{itemize}
 \item[(i)] The form factors $F_{0}$ and $A_{0,1,2}$ decrease with
 the increasing parameters ${\omega}$ and $v$. The form factors of
 $F_{0}$ and $A_{0,1}$ are more sensitive to the parameters ${\omega}$
 or/and $v$ than the form factor of $A_{2}$.
 Uncertainties of form factors $F_{0}$ and $A_{0,1}$ subjected to the
 charmonium wave function for Coulomb potential are larger than those
 for harmonic oscillator potential.
 Uncertainties of $F_{0}$ and $A_{0,1}$ related to the parameters
 ${\omega}$ in our given range are about $16\%$ ${\sim}$ $20\%$,
 while those related to the parameters $v$ in our given range are
 about $20\%$ ${\sim}$ $30\%$.
 In addition, the uncertainties of the decay constants $f_{B_{c}}$,
 $f_{J/{\psi}}$ and $f_{{\eta}_{c}}$ will bring ${\sim}$ $0.8\%$, $3\%$
 and $12\%$ uncertainties to the form factors $F_{0}$ and $A_{0,1,2}$,
 respectively.
 \item[(ii)]
 The form factors have been widely studied in the previous works
 \cite{prd39p1342,zpc57p43,prd48p5208,npb569p473,prd63p074010,
 prd68p094020,prd71p094006,prd74p074008,epjc51p833,epjc51p841,08113748}.
 There are large difference among the predictions in respect of
 various approaches.
 Compared with the previous results where $F_{0}^{B_{c}{\to}{\eta}_{c}}$
 ${\approx}$ $A_{0,1,2}^{B_{c}{\to}J/{\psi}}$ \cite{prd39p1342,
 npb569p473,prd63p074010,prd68p094020,prd71p094006,prd74p074008,
 epjc51p841,08113748}, our numerical results show that
 $F_{0}^{B_{c}{\to}{\eta}_{c}}$ ${\approx}$
 $A_{0}^{B_{c}{\to}J/{\psi}}$ $>$ $A_{1}^{B_{c}{\to}J/{\psi}}$
 $>$ $A_{2}^{B_{c}{\to}J/{\psi}}$.
 With some appropriate parameters, our results
 \footnote{For example,
 $F_{0}^{B_{c}{\to}{\eta}_{c}}$ $=$ $0.430$ ($0.464$),
 $A_{0}^{B_{c}{\to}J/{\psi}}$ $=$ $0.446$ ($0.470$),
 $A_{1}^{B_{c}{\to}J/{\psi}}$ $=$ $0.392$ ($0.427$)
 for $v$ $=$ $0.80$ (${\omega}$ $=$ $1.60$ GeV).}
 on the form factors $F_{0}$ and $A_{0,1}$ are in agreement with
 those in the previous works \cite{prd39p1342,npb569p473,
 prd63p074010,prd68p094020,prd71p094006,prd74p074008,
 epjc51p841,08113748}.
 Our results on the form factors $A_{2}$ are smaller than those
 in the previous works \cite{npb569p473,prd63p074010,
 prd68p094020,prd71p094006,prd74p074008,epjc51p841,08113748}.
 According to the ``spectator quark'' ansatz, there might be
 $F_{0}^{B_{c}{\to}{\eta}_{c}}$ (or $A_{0,1,2}^{B_{c}{\to}J/{\psi}}$)
 ${\sim}$ $F_{0}^{B{\to}D}$ (or $A_{0,1,2}^{B{\to}D^{\ast}}$)
 ${\approx}$ $0.6$ by intuition. So maybe the results based on the
 three-point QCD sum rules \cite{zpc57p43,prd48p5208} are small.
 \end{itemize}

\begin{figure}[ht]
 \includegraphics[250,685][350,770]{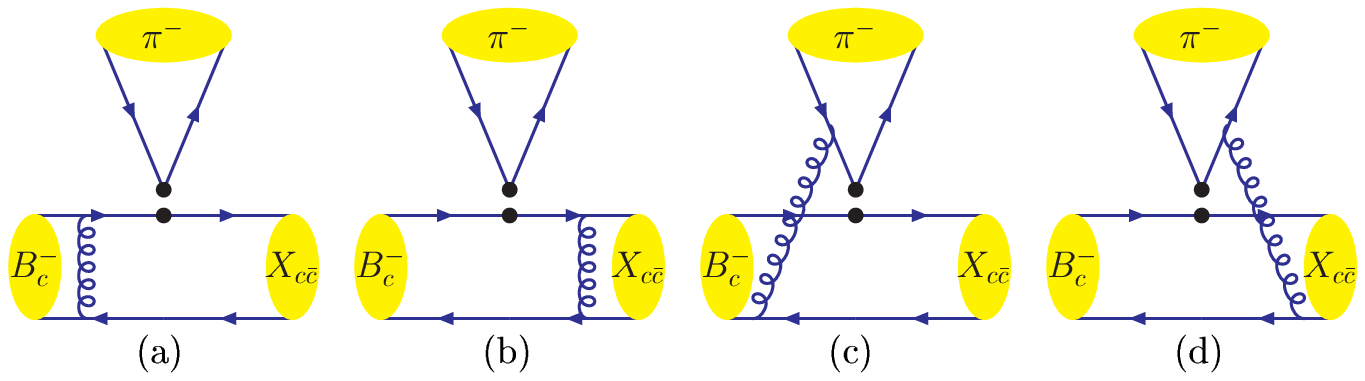}
 \caption{Feynman diagrams for $B_{c}^{-}$ ${\to}$ $X_{c\bar{c}}{\pi}^{-}$ decays.}
 \label{fig3}
 \end{figure}

 Our numerical results on the amplitudes and branching ratios for
 $B_{c}$ ${\to}$ $J/{\psi}{\pi}$, ${\eta}_{c}{\pi}$ decays are listed
 in TABLE.\ref{tab2} and \ref{tab3}.
 From the numbers in TABLE.\ref{tab2} and \ref{tab3}, we can see that
 \begin{itemize}
 \item[(i)] The contributions of the nonfactorizable topology
 [FIG.\ref{fig3} (c) and (d)] can provide large strong phases.
 The strong phases of the FIG.\ref{fig3} (c) topology ${\delta}$
 ${\gtrsim}$ $110^{\circ}$, while the strong phases of the
 FIG.\ref{fig3} (d) topology ${\delta}$ ${\lesssim}$ $-50^{\circ}$.
 The interferences between FIG.\ref{fig3} (c) and (d) are
 destructive.
 The strong phases from nonfactorizable topology decrease with
 the increasing parameters ${\omega}$ and $v$. They are free
 from the uncertainties of the decay constants $f_{B_{c}}$,
 $f_{J/{\psi}}$ and $f_{{\eta}_{c}}$. The strong phases
 subjected to the charmonium wave function for Coulomb potential
 are larger than those for harmonic oscillator potential in our
 given ranges.
 \item[(ii)] The dominated contributions to the branching ratios
 come from the factorizable topology [FIG.\ref{fig3} (a) and (b)].
 The ratio of amplitudes
 ${\vert}{\cal A}_{\rm FIG.\ref{fig3}(c+d)}/{\cal A}_{\rm FIG.\ref{fig3}(a+b)}{\vert}$
 ${\sim}$ $1\%$ for $B_{c}$ ${\to}$ $J/{\psi}{\pi}$ decay, and
 about $2\%$ ${\sim}$ $3 \%$ for $B_{c}$ ${\to}$ ${\eta}_{c}{\pi}$
 decay. The dominating amplitudes ${\cal A}_{\rm FIG.\ref{fig3}(a,b)}$
 and the branching ratios decrease with the increasing parameters
 ${\omega}$ and $v$. Besides the large uncertainties from the
 parameter ${\omega}$ and $v$, the uncertainties of the decay
 constants $f_{J/{\psi}}$ and $f_{{\eta}_{c}}$ will bring
 ${\sim}$ $7\%$ and ${\sim}$ $24\%$ uncertainties to the branching
 ratio for $B_{c}$ ${\to}$ $J/{\psi}{\pi}$ and ${\eta}_{c}{\pi}$
 decays, respectively.
 Considered the uncertainties from the input parameters,
 our results on the branching ratios are basically consistent with
 those in previous works
 \cite{epjc51p841,cpl18p498,prd49p3399,npb440p251,0211021,prd77p074013,prd73p054024}
 (see the numbers in TABLE.\ref{tab4}).
 Compared with the results in \cite{prd68p094020,prd74p074008}
 where small form factors are used (see the numbers in TABLE.\ref{tab1}),
 we find that our predictions are large. If with the same factor factors,
 our results generally agree with those in \cite{prd68p094020,prd74p074008}.
 The large predictions in \cite{prd65p114007} are obtained by the
 relations among the amplitudes under the quark diagram scheme, i.e.
 ${\cal A}(B_{c}{\to}{\eta}_{c}{\pi})$ $=$
 $V_{cb}/V_{ub}{\cal A}(B{\to}D_{s}{\pi})$ and
 ${\cal A}(B_{c}{\to}J/{\psi}{\pi})$ $=$
 $V_{cb}/V_{ub}{\cal A}(B{\to}D_{s}{\rho})$.
 Intuitively, the distribution amplitudes of the heavy quarkonia
 (such as $B_{c}$, $J/{\psi}$ and ${\eta}_{c}$ mesons) should be narrower
 than those of the ``heavy-light'' systems (such as $B$ and $D$
 mesons). So, the superposition among the $B_{c}-J/{\psi}({\eta}_{c})-{\pi}$
 systems might be less than those among the $B-D_{s}-{\pi}({\rho})$
 systems, i.e. there might be
 ${\cal A}(B_{c}{\to}{\eta}_{c}{\pi})$ ${\lesssim}$
 $V_{cb}/V_{ub}{\cal A}(B{\to}D_{s}{\pi})$ and
 ${\cal A}(B_{c}{\to}J/{\psi}{\pi})$ ${\lesssim}$
 $V_{cb}/V_{ub}{\cal A}(B{\to}D_{s}{\rho})$. If this argument or/and assumption
 is true, then it is expected that the results in \cite{prd65p114007} would
 become smaller and be consistent with ours.
 \item[(iii)] The dominating amplitudes ${\cal A}_{\rm FIG.\ref{fig3}(a,b)}$
 and the branching ratios subjected to the charmonium wave function
 for Coulomb potential are larger than those for harmonic oscillator
 potential. For a fixed value of parameter ${\omega}$ or $v$, the
 relation of the branching ratios is ${\cal BR}(B_{c}{\to}{\eta}_{c}{\pi})$
 ${\gtrsim}$ ${\cal BR}(B_{c}{\to}J/{\psi}{\pi})$. There are at
 least two reasons, one is that the phase spaces for $B_{c}$ ${\to}$
 ${\eta}_{c}{\pi}$ decay is larger than those for $B_{c}$ ${\to}$
 $J/{\psi}{\pi}$ decay, the other is that $F_{0}^{B_{c}{\to}{\eta}_{c}}$
 ${\gtrsim}$ $A_{0}^{B_{c}{\to}J/{\psi}}$ (see the numbers in TABLE.\ref{tab1}).
 The relation of the branching ratios is in agreement with the previous
 predictions \cite{prd39p1342,prd68p094020,prd74p074008,prd49p3399,npb440p251,
 prd65p114007,0211021,prd77p074013,prd73p054024}. The signal of $B_{c}$ ${\to}$
 $J/{\psi}{\pi}$ decay has been identified by the detectors at hadron collider
 Tevatron. It is eagerly expected that the signal of $B_{c}$ ${\to}$
 ${\eta}_{c}{\pi}$ decay is at the near corner for Tevatron and LHC.
 \end{itemize}

 \section{Summary and Conclusion}
 \label{sec4}
 In this paper, the $B_{c}$ ${\to}$ $J/{\psi}{\pi}$, ${\eta}_{c}{\pi}$
 decays are studied with the perturbative QCD approach. It is found that
 the form factors $A_{0,1,2}^{B_{c}{\to}J/{\psi}}$ and
 $F_{0}^{B_{c}{\to}{\eta}_{c}}$ for the $B_{c}$ ${\to}$ $J/{\psi}$,
 ${\eta}_{c}$ transitions and the branching ratios for $B_{c}$ ${\to}$
 $J/{\psi}{\pi}$ and ${\eta}_{c}{\pi}$ decays, they decrease with the
 increasing parameters ${\omega}$ and $v$, where ${\omega}$ and $v$
 are the parameters of the charmonium wave functions for Coulomb
 potential and harmonic oscillator potential, respectively. Therefore,
 the $B_{c}$ ${\to}$ $J/{\psi}{\pi}$, ${\eta}_{c}{\pi}$ decay modes
 provide good places to test quark potential models. In addition,
 the large uncertainties come from the uncertainties of the decay constants
 $f_{J/{\psi}}$ and $f_{{\eta}_{c}}$, which could be reduced greatly
 with the more accurate experimental measurements or/and better
 theoretical calculations. There are some other uncertainties not
 considered here, such as the power suppressed terms, the high order
 corrections, the effects of the final states interaction, the
 relativistic corrections to the wave functions, the model
 dependencies of the wave functions, and so on. They might be
 important in some cases (for example, the chirally-enhanced
 power corrections to the $B$ ${\to}$ ${\pi}K$ decays are not
 much suppressed numerically.) and deserve the dedicated researches.
 So our results might be regarded as the estimations under the
 pQCD framework. One should not be too serious about these numbers.
 Anyway, the large branching ratios and the clear signals of the final
 states make the measurement of the interesting $B_{c}$ ${\to}$
 $J/{\psi}{\pi}$, ${\eta}_{c}{\pi}$ decays easily at the hadron
 colliders.

 \begin{appendix}
 \section{Wave functions of the ${\pi}$ meson}
 \label{app01}
 The distribution amplitude ${\phi}_{\pi}^{a}$ for the twist-2
 wave function and the distribution amplitude ${\phi}_{\pi}^{p}$ and
 ${\phi}_{\pi}^{t}$ for the twist-3 wave functions are \cite{prd65p014007}
  \begin{eqnarray}
 {\phi}_{\pi}^{a}(x)&=&\frac{f_{\pi}}{2\sqrt{2N_{c}}}6x\bar{x} \Big\{
          1+0.44C_{2}^{3/2}(\bar{x}-x)+0.25C_{4}^{3/2}(\bar{x}-x)\Big\}
  \label{eq:pi-a}, \\
 {\phi}_{\pi}^{p}(x)&=&\frac{f_{\pi}}{2\sqrt{2N_{c}}}\Big\{
          1+0.43C_{2}^{1/2}(\bar{x}-x)+0.09C_{4}^{1/2}(\bar{x}-x)\Big\}
  \label{eq:pi-p}, \\
 {\phi}_{\pi}^{t}(x)&=&\frac{f_{\pi}}{2\sqrt{2N_{c}}}\Big\{C_{1}^{1/2}(\bar{x}-x)
          +0.55C_{3}^{1/2}(\bar{x}-x)\Big\}
  \label{eq:pi-t},
  \end{eqnarray}
  with the decay constant $f_{\pi}$ $=$ $130$ MeV.
  The Gegenbauer polynomials are defined by
  \begin{eqnarray}
   &&C_{1}^{1/2}(z)=z,~~~~~~~~~~~~~~~~~~~
     C_{2}^{1/2}(z)=\frac{1}{2}(3z^{2}-1), \nonumber \\
   &&C_{3}^{1/2}(z)=\frac{1}{2}(5z^{3}-3z),~~~~~
     C_{4}^{1/2}(z)=\frac{1}{8}(35z^{4}-30z^{2}+3), \nonumber \\
   &&C_{2}^{3/2}(z)=\frac{3}{2}(5z^{2}-1),~~~~~~~
     C_{4}^{3/2}(z)=\frac{15}{8}(21z^{4}-14z^{2}+1). \nonumber
  \end{eqnarray}

 \section{Wave functions of the $J/{\psi}$ and ${\eta}_{c}$ mesons}
 \label{app02}
 The heavy quarkonium, such as $c\bar{c}$, similar to diatomic molecules,
 might be amenable to a Born-Oppenheimer treatment
 \footnote{The heavy quark-antiquark pair is bound by the gluon and light-quark
 clouds. The heavy quarks correspond to the nuclei in diatomic molecules.
 The gluon and light-quark fields correspond to the electrons, and provide
 adiabatic potentials \cite{0412158}.} \cite{0412158}.
 Following the prescription in \cite{prd71p114008,plb612p215}, two forms
 of the wave functions corresponding to two different nonrelativistic
 potentials will be derived.
 \subsection{wave functions for harmonic oscillator potential}
 \label{app02a}
 In the nuclear shell model, a more realistic description of the nucleons
 inside the atomic nucleus is given by the Woods-Saxon potential.
 The Schr\"{o}dinger equation subjected to the Woods-Saxon potential cannot
 be solved analytically, and must be treated numerically, but the energy
 levels as well as other properties can be arrived at by approximating the
 model with a three-dimensional harmonic oscillator. The spectroscopy of the
 heavy quarkonium $c\bar{c}$ can be treated by this model.
 The quantum number $n_{r}L$ for the $J/{\psi}$ and ${\eta}_{c}$ mesons is $1S$,
 where $n_{r}$ and $L$ are the radial quantum number and the orbital angular
 momentum quantum number, respectively. (note : the energy spectrum of
 a three-dimensional harmonic oscillator is given by
 $E_{n_{r}L}$ $=$ $\{ 2(n_{r}-1)+L+\frac{3}{2} \}{\omega}$.)
 The radial wave function of the corresponding Schr\"{o}dinger state is given by
 \begin{equation}
 {\psi}_{n_{r}L}(r)={\psi}_{1S}(r)~{\propto}~{\exp}(-{\alpha}^{2}r^{2}/2)
 \label{eq:oscillator-01},
 \end{equation}
 where ${\alpha}^{2}$ $=$ $m_{c}{\omega}/2$, ${\omega}$ is the
 frequency of oscillations or the quantum of energy.

 Applying the Fourier transform, the state Eq.(\ref{eq:oscillator-01}) is
 replaced by the mapping representation on the momentum space,
 \begin{equation}
 {\psi}_{1S}(\vec{k})~{\sim}~
 {\int}{\bf d}^{3}{\vec{r}}~e^{-i\vec{r}{\cdot}\vec{k}}{\psi}_{1S}(r)~
 {\propto}~{\exp}\Big(\frac{-\vec{k}^{2}}{2{\alpha}^{2}}\Big)
 \label{eq:oscillator-02}.
 \end{equation}

 Employing the substitution ansatz \cite{prd71p114008,plb612p215}:
 \begin{equation}
 \vec{k}_{\perp}{\to}\vec{k}_{\perp},~~~~~
 k_{z}{\to}(x-\bar{x})\frac{m_{0}}{2},~~~~~
 m_{0}^{2}=\frac{m_{c}^{2}+\vec{k}_{\perp}^{2}}{x\bar{x}},
 \label{eq:oscillator-03}
 \end{equation}
 where $\bar{x}$ $=$ $1$ $-$ $x$, and $x$ is the longitudinal momentum
 fraction of the valence quark of the meson, the wave function can be
 taken as
 \begin{equation}
 {\psi}_{1S}(\vec{k})~{\to}~{\psi}_{1S}(x,\vec{k}_{\perp})~{\propto}~
 {\exp}\Big(-\frac{(x-\bar{x})^{2}m_{c}^{2}+\vec{k}_{\perp}^{2}}
                  {8{\alpha}^{2}x\bar{x}}\Big)
 \label{eq:oscillator-04}.
 \end{equation}

 Applying the Fourier transform to replace the transverse momentum
 $\vec{k}_{\perp}$ with its conjugate variable $\vec{b}$,
 the $1S$-oscillator wave function can be taken as
  \begin{equation}
 {\psi}_{1S}(x,b)~{\sim}~
 {\int}{\bf d}^{2}\vec{k}_{\perp}e^{-i\vec{b}{\cdot}\vec{k}_{\perp}}
 {\psi}_{1S}(x,\vec{k}_{\perp})~{\propto}~ x\bar{x}{\exp}\Big\{
  -\frac{m_{c}}{\omega}x\bar{x}\Big[\Big(\frac{x-\bar{x}}{2x\bar{x}}\Big)^{2}
  +{\omega}^{2}b^{2} \Big] \Big\}
 \label{eq:oscillator-05}.
 \end{equation}

 The modified wave functions can be written as
 \begin{equation}
 {\psi}_{X_{c\bar{c}}(1S)}(x,b)~{\propto}~{\Phi}^{\rm asy}(x){\exp}\Big\{
  -\frac{m_{c}}{\omega}x\bar{x}\Big[\Big(\frac{x-\bar{x}}{2x\bar{x}}\Big)^{2}
  +{\omega}^{2}b^{2} \Big] \Big\}
 \label{eq:oscillator-06},
 \end{equation}
 with ${\Phi}^{\rm asy}(x)$ being set to the asymptotic models of the
 corresponding twists for light mesons, which have been given in \cite{plb612p215}.
 Therefore, we can obtain the wave functions of the $J/{\psi}$ and
 ${\eta}_{c}$ mesons in Eq.(\ref{eq:wf-jpsi-01}) and Eq.(\ref{eq:wf-etac-01})
 \begin{eqnarray}
 {\phi}_{\psi}^{L}(x,b) &=& \frac{f_{J/{\psi}}}{2\sqrt{2N_{c}}} N^{L}_{\psi}
 x\bar{x}{\exp}\Big\{-\frac{m_{c}}{\omega}x\bar{x}\Big[
  \Big(\frac{x-\bar{x}}{2x\bar{x}}\Big)^{2}
 +{\omega}^{2}b^{2} \Big] \Big\}
  \label{eq:wf-jpsi-02}, \\
 {\phi}_{\psi}^{t}(x,b) &=& \frac{f_{J/{\psi}}}{2\sqrt{2N_{c}}} N^{t}_{\psi}
 (x-\bar{x})^{2}{\exp}\Big\{-\frac{m_{c}}{\omega}x\bar{x}\Big[
  \Big(\frac{x-\bar{x}}{2x\bar{x}}\Big)^{2}
 +{\omega}^{2}b^{2} \Big] \Big\}
  \label{eq:wf-jpsi-03},  \\
 {\phi}_{{\eta}_{c}}^{v}(x,b) &=& \frac{f_{{\eta}_{c}}}{2\sqrt{2N_{c}}} N^{v}_{{\eta}_{c}}
 x\bar{x}{\exp}\Big\{-\frac{m_{c}}{\omega}x\bar{x}\Big[
  \Big(\frac{x-\bar{x}}{2x\bar{x}}\Big)^{2}
 +{\omega}^{2}b^{2} \Big] \Big\}
  \label{eq:wf-etac-02}, \\
 {\phi}_{{\eta}_{c}}^{s}(x,b) &=& \frac{f_{{\eta}_{c}}}{2\sqrt{2N_{c}}} N^{s}_{{\eta}_{c}}
 {\exp}\Big\{-\frac{m_{c}}{\omega}x\bar{x}\Big[
  \Big(\frac{x-\bar{x}}{2x\bar{x}}\Big)^{2}
 +{\omega}^{2}b^{2} \Big] \Big\}
  \label{eq:wf-etac-03},
 \end{eqnarray}
 where $N_{c}$ is the color number,
 $N^{L,t}_{\psi}$, $N^{v,s}_{{\eta}_{c}}$ are the normalization constants.
 All wave function in Eqs.(\ref{eq:wf-jpsi-02})-(\ref{eq:wf-etac-03})
 are symmetric under $x$ ${\leftrightarrow}$ $\bar{x}$ and normalized :
 \begin{eqnarray}
 {\int}_{0}^{1}{\bf d}x~{\phi}_{\psi}^{L,t}(x,0)&=&\frac{f_{J/{\psi}}}{2\sqrt{2N_{c}}}
  \label{eq:wf-jpsi-04},  \\
 {\int}_{0}^{1}{\bf d}x~{\phi}_{{\eta}_{c}}^{v,s}(x,0)&=&\frac{f_{{\eta}_{c}}}{2\sqrt{2N_{c}}}
  \label{eq:wf-etac-04}.
 \end{eqnarray}
 The parameter ${\omega}$ ${\approx}$ $m_{{\psi}(2S)}$ $-$ $m_{J/{\psi}(1S)}$
 ${\approx}$ $m_{{\eta}_{c}(2S)}$ $-$ $m_{{\eta}_{c}(1S)}$ ${\approx}$ $0.6$ GeV.

 \subsection{wave functions for Coulomb potential}
 \label{app02b}
 In the static QCD potential, the interactions between heavy quarkonium
 can be parameterized and well described by a funnel shape Coulomb plus
 linear potential.
 At short distances one-gluon-exchange leads to the Coulomb-like
 potential with a strength proportional to the QCD coupling constant
 ${\alpha}_{s}$ \cite{0412158}
 \begin{equation}
 V(r)=-C_{F}\frac{{\alpha}_{s}(r)}{r}
 \label{eq:Colulomb-01},
 \end{equation}
 where $C_{F}$ $=$ $4/3$ is the $SU(3)$ colour factor.

 The radial wave function of the corresponding Schr\"{o}dinger state is
 given by (note : the principle quantum number $n$ associated with
 Coulomb potential is given by $n$ $=$ $(n_{r}-1)+L+1$)
 \begin{equation}
 {\psi}_{n_{r}L}(r)={\psi}_{1S}(r)~{\propto}~{\exp}(-q_{B}r)
 \label{eq:Colulomb-02},
 \end{equation}
 where $q_{B}$ $=$ $C_{F}{\mu}_{c}{\alpha}_{s}$ is the Bohr momentum,
 ${\mu}_{c}$ $=$ $m_{c}/2$ is the reduced mass of the $c$-quark.
 Analogous to the treatment for the case of harmonic oscillator
 discussed above, we can get
 \begin{eqnarray}
 {\psi}_{1S}(\vec{k}) &{\propto}& \frac{1}{(\vec{k}^{2}+q_{B}^{2})^{2}}
  \label{eq:Colulomb-03}, \\
 {\psi}_{1S}(x,b) &{\propto}& \frac{(x\bar{x})^{2}m_{c}b}{\sqrt{1-4x\bar{x}(1-v^{2})}}
  K_{1}(m_{c}b\sqrt{1-4x\bar{x}(1-v^{2})})
  \label{eq:Colulomb-04},
 \end{eqnarray}
 where the typical velocity of the quarks in charmonium $v$ $=$ $q_{B}/m_{c}$
 $=$ $2{\alpha}_{s}/3$ ${\sim}$ $0.3$ \cite{nppsb140p440}.
 The wave functions of the $J/{\psi}$ and ${\eta}_{c}$ mesons
 can be written as
  \begin{eqnarray}
 {\phi}_{\psi}^{L}(x,b) &=& \frac{f_{J/{\psi}}}{2\sqrt{2N_{c}}} N^{L}_{\psi}
  \frac{(x\bar{x})^{2}m_{c}b}{\sqrt{1-4x\bar{x}(1-v^{2})}}
  K_{1}(m_{c}b\sqrt{1-4x\bar{x}(1-v^{2})})
  \label{eq:wf-jpsi-05}, \\
 {\phi}_{\psi}^{t}(x,b) &=& \frac{f_{J/{\psi}}}{2\sqrt{2N_{c}}} N^{t}_{\psi}
  \frac{(x-\bar{x})^{2}x\bar{x}m_{c}b}{\sqrt{1-4x\bar{x}(1-v^{2})}}
  K_{1}(m_{c}b\sqrt{1-4x\bar{x}(1-v^{2})})
  \label{eq:wf-jpsi-06}, \\
 {\phi}_{{\eta}_{c}}^{v}(x,b) &=& \frac{f_{{\eta}_{c}}}{2\sqrt{2N_{c}}} N^{v}_{{\eta}_{c}}
  \frac{(x\bar{x})^{2}m_{c}b}{\sqrt{1-4x\bar{x}(1-v^{2})}}
  K_{1}(m_{c}b\sqrt{1-4x\bar{x}(1-v^{2})})
  \label{eq:wf-etac-05}, \\
 {\phi}_{{\eta}_{c}}^{s}(x,b) &=& \frac{f_{{\eta}_{c}}}{2\sqrt{2N_{c}}} N^{s}_{{\eta}_{c}}
  \frac{x\bar{x}m_{c}b}{\sqrt{1-4x\bar{x}(1-v^{2})}}
  K_{1}(m_{c}b\sqrt{1-4x\bar{x}(1-v^{2})})
  \label{eq:wf-etac-06}.
 \end{eqnarray}
 The normalization conditions are the same as those of Eq.(\ref{eq:wf-jpsi-04})
 and Eq.(\ref{eq:wf-etac-04}).

  \section{Form factors in the perturbative QCD approach}
 \label{app03}
 \begin{eqnarray}
 \lefteqn{ F_{0}^{B_{c}{\to}{\eta}_{c}}=8{\pi}m_{B_{c}}^{2}C_{F}
 {\int}_{0}^{1}{\bf d}x_{1}{\bf d}x_{2}{\int}_{0}^{\infty}\!b_{2}{\bf d}b_{2}\,
 {\phi}_{B_{c}}(x_{1}) } \nonumber \\ &{\times}&
 \Big\{E_{a}(t_{a}){\alpha}_{s}(t_{a})H_{a}({\alpha},{\beta}_{a},b_{2})
 \Big[r_{{\eta}_{c}}\Big(2(1-x_{2})-r_{b}\Big){\phi}_{{\eta}_{c}}^{s}(x_{2},b_{2})
 \nonumber \\  &&~~~~~~~~~~~~~~~~~~~~~~~~~~~~~~~~~~~
 -\Big((1-x_{2})-2r_{b}\Big){\phi}_{{\eta}_{c}}^{v}(x_{2},b_{2}) \Big]
 \nonumber \\ &&
 -E_{b}(t_{b}){\alpha}_{s}(t_{b})H_{b}({\alpha},{\beta}_{b},b_{2})\Big[
 \Big(r_{{\eta}_{c}}^{2}(1-x_{1})+r_{c}\Big){\phi}_{{\eta}_{c}}^{v}(x_{2},b_{2})
 \nonumber \\  &&~~~~~~~~~~~~~~~~~~~~~~~~~~~~~
 -2r_{{\eta}_{c}}\Big((1-x_{1})+r_{c}\Big){\phi}_{{\eta}_{c}}^{s}(x_{2},b_{2})\Big]\Big\}
 \label{eq:formfactor-f0-01},
 \end{eqnarray}
 \begin{eqnarray}
 \lefteqn{ \frac{m_{B_{c}}+m_{J/{\psi}}}{2m_{J/{\psi}}}A_{1}^{B_{c}{\to}J/{\psi}} =
 -4{\pi}m_{B_{c}}^{2}C_{F}{\int}_{0}^{1}{\bf d}x_{1}{\bf d}x_{2}
 {\int}_{0}^{\infty}\!b_{2}{\bf d}b_{2}\,{\phi}_{B_{c}}(x_{1}) }
 \nonumber \\ &{\times}&
 \Big\{E_{a}(t_{a}){\alpha}_{s}(t_{a})H_{a}({\alpha},{\beta}_{a},b_{2})
 \Big[ \Big(2-x_{2}-4r_{b}-x_{2}r_{\psi}^{2}\Big){\phi}_{\psi}^{L}(x_{2},b_{2})
 \nonumber \\ &&~~~~~~~~~~~~~~~~
 + \Big(r_{b}r_{\psi}-2r_{\psi}+4x_{2}r_{\psi}
      +\frac{r_{b}}{r_{\psi}}-\frac{2}{r_{\psi}}\Big)
      {\phi}_{\psi}^{t}(x_{2},b_{2}) \Big]
 \nonumber \\ &&
  -E_{b}(t_{b}){\alpha}_{s}(t_{b})H_{b}({\alpha},{\beta}_{b},b_{2})
 \Big(1+2r_{c}-2x_{1}+r_{\psi}^{2}\Big){\phi}_{\psi}^{L}(x_{2},b_{2}) \Big\}
 \label{eq:formfactor-a1-01},
 \end{eqnarray}
 \begin{eqnarray}
 \lefteqn{ \frac{m_{B_{c}}-m_{J/{\psi}}}{-2m_{J/{\psi}}}A_{2}^{B_{c}{\to}J/{\psi}} =
 -4{\pi}m_{B_{c}}^{2}C_{F}{\int}_{0}^{1}{\bf d}x_{1}{\bf d}x_{2}
 {\int}_{0}^{\infty}\!b_{2}{\bf d}b_{2}\,{\phi}_{B_{c}}(x_{1}) }
 \nonumber \\ &{\times}&
 \Big\{E_{a}(t_{a}){\alpha}_{s}(t_{a})H_{a}({\alpha},{\beta}_{a},b_{2})
 \Big[ \Big(-x_{2}+x_{2}r_{\psi}^{2}\Big){\phi}_{\psi}^{L}(x_{2},b_{2})
 \nonumber \\ &&~~~~~~~~~~~~~~~~~~~~
 + \Big(r_{b}r_{\psi}-2r_{\psi}-\frac{r_{b}}{r_{\psi}}+\frac{2}{r_{\psi}}\Big)
      {\phi}_{\psi}^{t}(x_{2},b_{2}) \Big]
 \nonumber \\ &&
  +E_{b}(t_{b}){\alpha}_{s}(t_{b})H_{b}({\alpha},{\beta}_{b},b_{2})
  (1-2x_{1})(1-r_{\psi}^{2}){\phi}_{\psi}^{L}(x_{2},b_{2}) \Big\}
 \label{eq:formfactor-a2-01},
 \end{eqnarray}
 where $t_{a(b)}$ $=$ ${\max}(\sqrt{{\vert}{\alpha}{\vert}},
 \sqrt{{\vert}{\beta}_{a(b)}{\vert}},1/b_{2})$,
 $E_{a(b)}(t)$ $=$ $e^{-S_{\psi}(t)}$,
  \begin{eqnarray}
 {\alpha}&=&-m_{B_{c}}^{2}(x_{1}-x_{2})(x_{1}-r_{\psi}^{2}x_{2}),
  \label{eq:alpha} \\
 {\beta}_{a} &=&-m_{B_{c}}^{2}\big[(1-x_{2})(1-r_{\psi}^{2}x_{2})-r_{b}^{2}\big],
  \label{eq:beta-a}  \\
 {\beta}_{b} &=&-m_{B_{c}}^{2}\big[(1-x_{1})(r_{\psi}^{2}-x_{1})-r_{c}^{2}\big],
  \label{eq:beta-b} \\
 S_{\psi}(t)&=&s(x_{2}p_{2}^{+},b_{2})+s(\bar{x}_{2}p_{2}^{+},b_{2})
    +2{\int}_{1/b_{2}}^{t}\frac{{\bf d}{\mu}}{\mu}{\gamma}_{q}
  \label{eq:sudakov-jpsi}.
 \end{eqnarray}
 The quark anomalous dimension ${\gamma}_{q}$ $=$ $-{\alpha}_{s}/{\pi}$.
 The explicit expression of $s(Q,b)$ appearing in Sudakov form factor can be
 found in \cite{npb642p263}. The hard functions $H$ are
 \begin{eqnarray}
 H_{a}({\alpha},{\beta},b)&=&
       \frac{K_{0}(b\sqrt{\alpha})-K_{0}(b\sqrt{\beta})}{{\beta}-{\alpha}}
  \label{eq:denominator-a}, \\
 H_{b}({\alpha},{\beta},b)&=&\frac{K_{0}(b\sqrt{\alpha})}{\beta}
  \label{eq:denominator-b}.
 \end{eqnarray}

 \section{The decay amplitudes}
 \label{app04}
 \subsection{The amplitudes for $B_{c}$ ${\to}$ $J/{\psi}{\pi}$ decay with
 the perturbative QCD approach}
 \label{app04a}
 \begin{eqnarray}
 {\cal A}_{\rm FIG.\ref{fig3}(a)}&=&8{\pi}C_{F}f_{\pi}
 m_{B_{c}}^{4}(1-r_{\psi}^{2})
 {\int}_{0}^{1}{\bf d}x_{1}{\bf d}x_{2}
 {\int}_{0}^{\infty}\!b_{2}{\bf d}b_{2}\,
 {\phi}_{B_{c}}(x_{1})E_{a}(t_{a}) \nonumber \\ &{\times}&
 {\alpha}_{s}(t_{a})C_{a}(t_{a})H_{a}({\alpha},{\beta}_{a},b_{2})
  \Big\{r_{\psi}\big[2(1-x_{2})-r_{b}\big]{\phi}_{\psi}^{t}(x_{2},b_{2})
  \nonumber \\ & &~~~~~~~~~~~~~~~~~~~~~~~~~~~~~~~~~
    -\big[(1-x_{2})-2r_{b}\big]{\phi}_{\psi}^{L}(x_{2},b_{2}) \Big\}
 \label{eq:am-jpsi-a}, \\
 {\cal A}_{\rm FIG.\ref{fig3}(b)}&=&8{\pi}C_{F}f_{\pi}
 m_{B_{c}}^{4}(1-r_{\psi}^{2})
 {\int}_{0}^{1}{\bf d}x_{1}{\bf d}x_{2}
 {\int}_{0}^{\infty}\!b_{2}{\bf d}b_{2}\,
 {\phi}_{B_{c}}(x_{1})E_{b}(t_{b}) \nonumber \\ &{\times}&
 {\alpha}_{s}(t_{b})C_{b}(t_{b})H_{b}({\alpha},{\beta}_{b},b_{2})
 \Big\{r_{\psi}^{2}(1-x_{1})+r_{c}\Big\}{\phi}_{\psi}^{L}(x_{2},b_{2})
 \label{eq:am-jpsi-b}, \\
 {\cal A}_{\rm FIG.\ref{fig3}(c)}&=&\frac{32{\pi}C_{F}}{\sqrt{2N_{c}}}
 m_{B_{c}}^{4}(1-r_{\psi}^{2})
 {\int}_{0}^{1}{\bf d}x_{1}{\bf d}x_{2}{\bf d}x_{3}
 {\int}_{0}^{\infty}\!b_{2}{\bf d}b_{2}\,b_{3}{\bf d}b_{3}\,
 {\phi}_{B_{c}}(x_{1}){\phi}_{\pi}^{a}(x_{3})
 \nonumber \\ &{\times}&
 E_{c}(t_{c}){\alpha}_{s}(t_{c})C_{c}(t_{c})
 H_{c}({\alpha},{\beta}_{c},b_{2},b_{3})
 \Big\{ r_{\psi}(x_{1}-x_{2}){\phi}_{\psi}^{t}(x_{2},b_{2})
 \nonumber \\ &&~~~~~~~~~~~~~~~~~~~~~~~~~~~~~~~~~
 +(1-r_{\psi}^{2})(x_{1}-x_{3}){\phi}_{\psi}^{L}(x_{2},b_{2})\Big\}
 \label{eq:am-jpsi-c}, \\
 {\cal A}_{\rm FIG.\ref{fig3}(d)}&=&\frac{32{\pi}C_{F}}{\sqrt{2N_{c}}}
 m_{B_{c}}^{4}(1-r_{\psi}^{2})
 {\int}_{0}^{1}{\bf d}x_{1}{\bf d}x_{2}{\bf d}x_{3}
 {\int}_{0}^{\infty}\!b_{2}{\bf d}b_{2}\,b_{3}{\bf d}b_{3}\,
 {\phi}_{B_{c}}(x_{1}){\phi}_{\pi}^{a}(x_{3})
 \nonumber \\ &{\times}&
 E_{d}(t_{d}){\alpha}_{s}(t_{d}) C_{d}(t_{d})
 H_{d}({\alpha},{\beta}_{d},b_{2},b_{3})
 \Big\{ r_{\psi}(x_{1}-x_{2}){\phi}_{\psi}^{t}(x_{2},b_{2})
 \nonumber \\ &&~~~~~~~~~~~~~~~~
 +\big[2(x_{2}-x_{1})-(x_{2}-\bar{x}_{3})(1-r_{\psi}^{2})\big]
 {\phi}_{\psi}^{L}(x_{2},b_{2}) \Big\}
 \label{eq:am-jpsi-d},
 \end{eqnarray}
 where $t_{c(d)}$ $=$ ${\max}(\sqrt{{\vert}{\alpha}{\vert}},
 \sqrt{{\vert}{\beta}_{c(d)}{\vert}},1/b_{2},1/b_{3})$,
 $E_{c(d)}(t)$ $=$ $e^{-S_{\psi}(t)-S_{\pi}(t)}$,
 \begin{eqnarray}
 {\beta}_{c} &=&-m_{B_{c}}^{2}(x_{2}-x_{1})
  \big[(x_{2}-x_{1})-(x_{2}-x_{3})(1-r_{\psi}^{2})\big]
  \label{eq:beta-c},  \\
 {\beta}_{d} &=&-m_{B_{c}}^{2}(x_{2}-x_{1})
  \big[(x_{2}-x_{1})-(x_{2}-\bar{x}_{3})(1-r_{\psi}^{2})\big]
  \label{eq:beta-d},  \\
 S_{\pi}(t)&=&s(x_{3}p_{3}^{-},b_{3})+s(\bar{x}_{3}p_{2}^{-},b_{3})
    +2{\int}_{1/b_{3}}^{t}\frac{{\bf d}{\mu}}{\mu}{\gamma}_{q}
  \label{eq:sudakov-pi},
 \end{eqnarray}
 \begin{equation}
  C_{a(b)}=C_{1}+C_{2}/N_{c},~~~~~~~~~~
  C_{c(d)}=C_{2},
 \end{equation}
 \begin{eqnarray}
 H_{c(d)}({\alpha},{\beta},b_{2},b_{3})&=& \Big\{
   {\theta}(b_{2}-b_{3})K_{0}(b_{2}\sqrt{\alpha})I_{0}(b_{3}\sqrt{\alpha})
   +(b_{2}{\leftrightarrow}b_{3}) \Big\} \nonumber \\ &{\times}&
   \Big\{ {\theta}(+{\beta})K_{0}(b_{3}\sqrt{\beta})
   +\frac{i{\pi}}{2}{\theta}(-{\beta})H_{0}^{(1)}(b_{3}\sqrt{-{\beta}}) \Big\}
  \label{eq:denominator-c},
 \end{eqnarray}
 where $C_{1,2}$ are the Wilson coefficients. The definitions of other
 parameters are the same as those in APPENDIX \ref{app03}.

 \subsection{The amplitudes for $B_{c}$ ${\to}$ ${\eta}_{c}{\pi}$ decay with
 the perturbative QCD approach}
 \label{app04b}
 \begin{eqnarray}
 {\cal A}_{\rm FIG.\ref{fig3}(a)}&=&+i8{\pi}C_{F}f_{\pi}
 m_{B_{c}}^{4}(1-r_{{\eta}_{c}}^{2})
 {\int}_{0}^{1}{\bf d}x_{1}{\bf d}x_{2}
 {\int}_{0}^{\infty}\!b_{2}{\bf d}b_{2}\,
 {\phi}_{B_{c}}(x_{1})E_{a}(t_{a}) \nonumber \\ &{\times}&
 {\alpha}_{s}(t_{a})C_{a}(t_{a})H_{a}({\alpha},{\beta}_{a},b_{2})
  \Big\{r_{{\eta}_{c}}\big[2(1-x_{2})-r_{b}\big]
 {\phi}_{{\eta}_{c}}^{s}(x_{2},b_{2})
  \nonumber \\ &&~~~~~~~~~~~~~~~~~~~~~~~~~~~~~~~~~~
 -\big[(1-x_{2})-2r_{b}\big]{\phi}_{{\eta}_{c}}^{v}(x_{2},b_{2}) \Big\}
 \label{eq:am-etac-a}, \\
 {\cal A}_{\rm FIG.\ref{fig3}(b)}&=&-i8{\pi}C_{F}f_{\pi}
 m_{B_{c}}^{4}(1-r_{{\eta}_{c}}^{2})
 {\int}_{0}^{1}{\bf d}x_{1}{\bf d}x_{2}
 {\int}_{0}^{\infty}\!b_{2}{\bf d}b_{2}\,
 {\phi}_{B_{c}}(x_{1})E_{b}(t_{b}) \nonumber \\ &{\times}&
 {\alpha}_{s}(t_{b}) C_{b}(t_{b})H_{b}({\alpha},{\beta}_{b},b_{2})
  \Big\{\big[r_{{\eta}_{c}}^{2}(1-x_{1})+r_{c}\big]
 {\phi}_{{\eta}_{c}}^{v}(x_{2},b_{2})
  \nonumber \\ &&~~~~~~~~~~~~~~~~~~~~~~~~~~~
 -2r_{{\eta}_{c}}\big[(1-x_{1})+r_{c}\big]
 {\phi}_{{\eta}_{c}}^{s}(x_{2},b_{2}) \Big\}
 \label{eq:am-etac-b}, \\
 {\cal A}_{\rm FIG.\ref{fig3}(c)}&=&\frac{-i32{\pi}C_{F}}{\sqrt{2N_{c}}}
 m_{B_{c}}^{4}(1-r_{{\eta}_{c}}^{2})
 {\int}_{0}^{1}{\bf d}x_{1}{\bf d}x_{2}{\bf d}x_{3}
 {\int}_{0}^{\infty}\!b_{2}{\bf d}b_{2}\,b_{3}{\bf d}b_{3}\,
 {\phi}_{B_{c}}(x_{1}){\phi}_{\pi}^{a}(x_{3})
 \nonumber \\ &{\times}&
 E_{c}(t_{c}){\alpha}_{s}(t_{c})C_{c}(t_{c})
 H_{c}({\alpha},{\beta}_{c},b_{2},b_{3})
 \Big\{ r_{{\eta}_{c}}(x_{1}-x_{2}){\phi}_{{\eta}_{c}}^{s}(x_{2},b_{2})
 \nonumber \\ &&~~~~~~~~~~
 -\big[(1-r_{{\eta}_{c}}^{2})(x_{1}-x_{3})
 +2r_{{\eta}_{c}}^{2}(x_{1}-x_{2})\big]
 {\phi}_{{\eta}_{c}}^{v}(x_{2},b_{2}) \Big\}
 \label{eq:am-etac-c}, \\
 {\cal A}_{\rm FIG.\ref{fig3}(d)}&=&\frac{+i32{\pi}C_{F}}{\sqrt{2N_{c}}}
 m_{B_{c}}^{4}(1-r_{{\eta}_{c}}^{2})
 {\int}_{0}^{1}{\bf d}x_{1}{\bf d}x_{2}{\bf d}x_{3}
 {\int}_{0}^{\infty}\!b_{2}{\bf d}b_{2}\,b_{3}{\bf d}b_{3}\,
 {\phi}_{B_{c}}(x_{1}){\phi}_{\pi}^{a}(x_{3})
 \nonumber \\ &{\times}&
 E_{d}(t_{d}){\alpha}_{s}(t_{d})C_{d}(t_{d})
 H_{d}({\alpha},{\beta}_{d},b_{2},b_{3})
 \Big\{ r_{{\eta}_{c}}(x_{1}-x_{2}){\phi}_{{\eta}_{c}}^{s}(x_{2},b_{2})
 \nonumber \\ &&~~~~~~~~~~~~~~~
 +\big[2(x_{2}-x_{1})-(x_{2}-\bar{x}_{3})(1-r_{{\eta}_{c}}^{2})\big]
 {\phi}_{{\eta}_{c}}^{v}(x_{2},b_{2}) \Big\}
 \label{eq:am-etac-d}.
 \end{eqnarray}

 \end{appendix}

 \section*{Acknowledgments}
 This work is supported in part both by National Natural Science Foundation
 of China (under Grant No. 10805014, 10647119, 10710146, and 90403024) and by
 Natural Science Foundation of Henan Province, China. We would like to thank
 Prof. Deshan Yang, Dr. Xianqiao Yu, Dr. Yumin Wang, and Dr. Wei Wang for
 valuable discussions. We thanks the referees for their helpful comments.
 Junfeng Sun would like to thank the {\em Kavli Institute for Theoretical
 Physics, China} (KITP) for their hospitality while this work was started.

 \begin{table}[htb]
 \caption{Form factors of $F_{0}^{B_{c}{\to}{\eta}_{c}}$ and
 $A_{0,1,2}^{B_{c}{\to}J/{\psi}}$}
 \label{tab1}
 \begin{ruledtabular}
 \begin{tabular}{c|cccc}
 & $F_{0}^{B_{c}{\to}{\eta}_{c}}$
 & $A_{0}^{B_{c}{\to}J/{\psi}}$
 & $A_{1}^{B_{c}{\to}J/{\psi}}$
 & $A_{2}^{B_{c}{\to}J/{\psi}}$ \\ \hline
   ${\omega}$ $=$ $0.5$ GeV
 & $0.790$
 & $0.775$
 & $0.671$
 & $0.469$ \\
   ${\omega}$ $=$ $0.6$ GeV
 & $0.741$
 & $0.730$
 & $0.636$
 & $0.454$ \\
   ${\omega}$ $=$ $0.7$ GeV
 & $0.698$
 & $0.690$
 & $0.605$
 & $0.440$ \\
   ${\omega}$ $=$ $0.8$ GeV
 & $0.660$
 & $0.655$
 & $0.578$
 & $0.427$ \\ \hline
   $v$ $=$ $0.25$
 & $0.903$
 & $0.891$
 & $0.712$
 & $0.363$ \\
   $v$ $=$ $0.30$
 & $0.824$
 & $0.819$
 & $0.664$
 & $0.364$ \\
   $v$ $=$ $0.35$
 & $0.760$
 & $0.759$
 & $0.624$
 & $0.361$ \\
   $v$ $=$ $0.40$
 & $0.705$
 & $0.708$
 & $0.589$
 & $0.356$ \\ \hline
   \cite{prd39p1342}
 & $0.170$ ${\sim}$ $0.687$
 & $0.156$ ${\sim}$ $0.684$
 & $0.156$ ${\sim}$ $0.745$
 & $0.156$ ${\sim}$ $0.862$ \\
   \cite{zpc57p43}
 & $0.20{\pm}0.02$
 & $0.26{\pm}0.07$
 & $0.27{\pm}0.03$
 & $0.28{\pm}0.09$ \\
   \cite{prd48p5208}
 & $0.23{\pm}0.01$
 & $0.21{\pm}0.03$
 & $0.21{\pm}0.02$
 & $0.22{\pm}0.02$ \\
   \cite{npb569p473}
 & $0.66$
 & $0.60$
 & $0.63$
 & $0.69$ \\
   \cite{prd63p074010}
 & $0.76$
 & $0.69$
 & $0.68$
 & $0.66$ \\
  \cite{prd68p094020}
 & $0.47$
 & $0.40$
 & $0.50$
 & $0.73$ \\
  \cite{prd71p094006}
 & $0.61$
 & $0.56$
 & $0.56$
 & $0.54$ \\
  \cite{prd74p074008}
 & $0.49$
 & $0.45$
 & $0.49$
 & $0.56$ \\
  \cite{epjc51p833}
 & $0.87$
 & $0.27$
 & $0.75$
 & $1.69$ \\
  \cite{epjc51p841}
 & ------
 & $0.57^{+0.01}_{-0.02}$
 & $0.55^{+0.01}_{-0.03}$
 & $0.51^{+0.03}_{-0.04}$ \\
  \cite{08113748}
 & $0.61^{+0.01+0.03}_{-0.02-0.04}$
 & $0.53{\pm}0.01$
 & $0.50^{+0.01}_{-0.02}$
 & $0.44^{+0.02}_{-0.03}$
 \end{tabular}
 \end{ruledtabular}
 \end{table}

 \begin{table}[htb]
 \caption{Amplitudes and branching ratio for $B_{c}$ ${\to}$ $J/{\psi}{\pi}$
 decay, where ${\delta}$ is the strong phase.}
 \label{tab2}
 \begin{ruledtabular}
 \begin{tabular}{c|cccc|c}
 & ${\cal A}_{\rm FIG.\ref{fig3}(a)}$
 & ${\cal A}_{\rm FIG.\ref{fig3}(b)}$
 & ${\cal A}_{\rm FIG.\ref{fig3}(c)}$ [${\delta}$]
 & ${\cal A}_{\rm FIG.\ref{fig3}(d)}$ [${\delta}$]
 & ${\cal BR}(B_{c}{\to}J/{\psi}{\pi})$ \\ \hline
   ${\omega}$ $=$ $0.5$ GeV
 & $1.359$
 & $1.831$
 & $-0.115+i0.269$ [$113^{\circ}$]
 & $+0.132-i0.285$ [$-65^{\circ}$]
 & $1.913{\times}10^{-3}$ \\
   ${\omega}$ $=$ $0.6$ GeV
 & $1.235$
 & $1.767$
 & $-0.103+i0.267$ [$111^{\circ}$]
 & $+0.115-i0.285$ [$-68^{\circ}$]
 & $1.689{\times}10^{-3}$ \\
   ${\omega}$ $=$ $0.7$ GeV
 & $1.133$
 & $1.704$
 & $-0.093+i0.261$ [$110^{\circ}$]
 & $+0.101-i0.279$ [$-70^{\circ}$]
 & $1.506{\times}10^{-3}$ \\
   ${\omega}$ $=$ $0.8$ GeV
 & $1.049$
 & $1.643$
 & $-0.086+i0.253$ [$109^{\circ}$]
 & $+0.090-i0.271$ [$-72^{\circ}$]
 & $1.352{\times}10^{-3}$ \\ \hline
   $v$ $=$ $0.25$
 & $1.941$
 & $1.738$
 & $-0.140+i0.150$ [$133^{\circ}$]
 & $+0.157-i0.192$ [$-51^{\circ}$]
 & $2.542{\times}10^{-3}$ \\
   $v$ $=$ $0.30$
 & $1.692$
 & $1.686$
 & $-0.130+i0.162$ [$129^{\circ}$]
 & $+0.144-i0.203$ [$-55^{\circ}$]
 & $2.140{\times}10^{-3}$ \\
   $v$ $=$ $0.35$
 & $1.498$
 & $1.632$
 & $-0.121+i0.169$ [$126^{\circ}$]
 & $+0.130-i0.209$ [$-58^{\circ}$]
 & $1.834{\times}10^{-3}$ \\
   $v$ $=$ $0.40$
 & $1.342$
 & $1.576$
 & $-0.111+i0.174$ [$123^{\circ}$]
 & $+0.118-i0.212$ [$-61^{\circ}$]
 & $1.591{\times}10^{-3}$
 \end{tabular}
 \end{ruledtabular}
 \end{table}

 \begin{table}[htb]
 \caption{Amplitudes and branching ratio for $B_{c}$ ${\to}$ ${\eta}_{c}{\pi}$
 decay, where ${\delta}$ is the strong phase.}
 \label{tab3}
 \begin{ruledtabular}
 \begin{tabular}{c|cccc|c}
 & $-i{\cal A}_{\rm FIG.\ref{fig3}(a)}$
 & $-i{\cal A}_{\rm FIG.\ref{fig3}(b)}$
 & $-i{\cal A}_{\rm FIG.\ref{fig3}(c)}$ [${\delta}$]
 & $-i{\cal A}_{\rm FIG.\ref{fig3}(d)}$ [${\delta}$]
 & ${\cal BR}(B_{c}{\to}{\eta}_{c}{\pi})$ \\ \hline
   ${\omega}$ $=$ $0.5$ GeV
 & $1.490$
 & $1.828$
 & $-0.112+i0.211$ [$118^{\circ}$]
 & $+0.127-i0.300$ [$-67^{\circ}$]
 & $2.117{\times}10^{-3}$ \\
   ${\omega}$ $=$ $0.6$ GeV
 & $1.353$
 & $1.756$
 & $-0.098+i0.214$ [$115^{\circ}$]
 & $+0.112-i0.299$ [$-69^{\circ}$]
 & $1.858{\times}10^{-3}$ \\
   ${\omega}$ $=$ $0.7$ GeV
 & $1.242$
 & $1.685$
 & $-0.087+i0.212$ [$112^{\circ}$]
 & $+0.100-i0.292$ [$-71^{\circ}$]
 & $1.646{\times}10^{-3}$ \\
   ${\omega}$ $=$ $0.8$ GeV
 & $1.149$
 & $1.617$
 & $-0.078+i0.207$ [$111^{\circ}$]
 & $+0.089-i0.283$ [$-73^{\circ}$]
 & $1.470{\times}10^{-3}$ \\ \hline
   $v$ $=$ $0.25$
 & $2.140$
 & $1.665$
 & $-0.131+i0.104$ [$141^{\circ}$]
 & $+0.154-i0.189$ [$-51^{\circ}$]
 & $2.792{\times}10^{-3}$ \\
   $v$ $=$ $0.30$
 & $1.864$
 & $1.608$
 & $-0.121+i0.119$ [$135^{\circ}$]
 & $+0.141-i0.201$ [$-55^{\circ}$]
 & $2.323{\times}10^{-3}$ \\
   $v$ $=$ $0.35$
 & $1.649$
 & $1.548$
 & $-0.112+i0.129$ [$131^{\circ}$]
 & $+0.129-i0.208$ [$-58^{\circ}$]
 & $1.970{\times}10^{-3}$ \\
   $v$ $=$ $0.40$
 & $1.476$
 & $1.488$
 & $-0.102+i0.137$ [$127^{\circ}$]
 & $+0.118-i0.212$ [$-61^{\circ}$]
 & $1.692{\times}10^{-3}$
 \end{tabular}
 \end{ruledtabular}
 \end{table}

 \begin{table}[htb]
 \caption{Branching ratio for $B_{c}$ ${\to}$ $J/{\psi}{\pi}$,
  ${\eta}_{c}{\pi}$ decay in previous works (in the unit of $10^{-3}$).}
 \label{tab4}
 \begin{ruledtabular}
 \begin{tabular}{c|rrrrrr}
   ${\cal BR}(B_{c}{\to}{\eta}_{c}{\pi})$
 & $0.13{\sim}1.55$ \cite{prd39p1342}
 & $0.85$ \cite{prd68p094020}
 & $0.94$ \cite{prd74p074008}
 & $1.44{\sim}2.46$ \cite{cpl18p498}
 & $2.30$ \cite{prd49p3399}
 & $1.8$ \cite{npb440p251} \\
 & $0.26$ \cite{prd61p034012}
 & $9.30$ \cite{prd65p114007}
 & $2.00$ \cite{0211021}
 & $1.16{\sim}1.34$ \cite{prd77p074013}
 & $1.90$ \cite{prd73p054024} \\ \hline
   ${\cal BR}(B_{c}{\to}J/{\psi}{\pi})$
 & $0.02{\sim}0.34$ \cite{prd39p1342}
 & $0.61$ \cite{prd68p094020}
 & $0.76$ \cite{prd74p074008}
 & $2.0^{+0.8+0.0}_{-0.7-0.1}$ \cite{epjc51p841}
 & $2.19$ \cite{prd49p3399}
 & $1.7$ \cite{npb440p251} \\
 & $1.30$ \cite{prd61p034012}
 & $4.50$ \cite{prd65p114007}
 & $1.30$ \cite{0211021}
 & $1.08{\sim}1.24$ \cite{prd77p074013}
 & $1.70$ \cite{prd73p054024}
 \end{tabular}
 \end{ruledtabular}
 \end{table}

 \end{document}